\documentclass[reprint,twocolumn,aps,prc,a4paper,superscriptaddress,showpacs,preprintnumbers,amsmath,amssymb]{revtex4}
\usepackage{amssymb}
\usepackage{amsmath}
\usepackage{graphicx}
\usepackage{lscape}
\usepackage{booktabs,longtable}
\usepackage{mathptmx, courier, pifont}
\usepackage[scaled=0.92]{helvet}
\usepackage[T1]{fontenc}
\usepackage{textcomp}

\begin{document}

\title{First application of combined isochronous and Schottky mass spectrometry: Half-lives of fully-ionized $^{49}$Cr$^{24+}$ and $^{53}$Fe$^{26+}$ atoms}
\author{X.~L.~Tu}
\affiliation{Institute of Modern Physics, Chinese Academy of Sciences, Lanzhou 730000, People's Republic of China}
\affiliation{Max-Planck-Institut f\"{u}r Kernphysik, Saupfercheckweg 1, 69117 Heidelberg, Germany}

\author{X.~C.~Chen}
\affiliation{Institute of Modern Physics, Chinese Academy of Sciences, Lanzhou 730000, People's Republic of China}

\author{J.~T.~Zhang}
\affiliation{Institute of Modern Physics, Chinese Academy of Sciences, Lanzhou 730000, People's Republic of China}
\affiliation{School of Nuclear Science and Technology, Lanzhou University, Lanzhou 730000, People's Republic of China}

\author{P.~Shuai}
\affiliation{Institute of Modern Physics, Chinese Academy of Sciences, Lanzhou 730000, People's Republic of China}

\author{K.~Yue}
\affiliation{Institute of Modern Physics, Chinese Academy of Sciences, Lanzhou 730000, People's Republic of China}

\author{X.~Xu}
\affiliation{Institute of Modern Physics, Chinese Academy of Sciences, Lanzhou 730000, People's Republic of China}

\author{C.~Y.~Fu}
\affiliation{Institute of Modern Physics, Chinese Academy of Sciences, Lanzhou 730000, People's Republic of China}

\author{Q.~Zeng}
\affiliation{School of Nuclear Science and Engineering, East China University of Technology, NanChang 330013, People's Republic of China}

\author{X.~Zhou}
\affiliation{Institute of Modern Physics, Chinese Academy of Sciences, Lanzhou 730000, People's Republic of China}

\author{Y.~M.~Xing}
\affiliation{Institute of Modern Physics, Chinese Academy of Sciences, Lanzhou 730000, People's Republic of China}

\author{J.~X.~Wu}
\affiliation{Institute of Modern Physics, Chinese Academy of Sciences, Lanzhou 730000, People's Republic of China}

\author{R.~S.~Mao}
\affiliation{Institute of Modern Physics, Chinese Academy of Sciences, Lanzhou 730000, People's Republic of China}

\author{L.~J.~Mao}
\affiliation{Institute of Modern Physics, Chinese Academy of Sciences, Lanzhou 730000, People's Republic of China}

\author{K.~H.~Fang}
\affiliation{School of Nuclear Science and Technology, Lanzhou University, Lanzhou 730000, People's Republic of China}

\author{Z.~Y.~Sun}
\affiliation{Institute of Modern Physics, Chinese Academy of Sciences, Lanzhou 730000, People's Republic of China}

\author{M.~Wang}
\affiliation{Institute of Modern Physics, Chinese Academy of Sciences, Lanzhou 730000, People's Republic of China}

\author{J.~C.~Yang}
\affiliation{Institute of Modern Physics, Chinese Academy of Sciences, Lanzhou 730000, People's Republic of China}

\author{Yu.~A. Litvinov}
\affiliation{Institute of Modern Physics, Chinese Academy of Sciences, Lanzhou 730000, People's Republic of China}
\affiliation{GSI Helmholtzzentrum f\"{u}r Schwerionenforschung, Planckstra{\ss}e 1, 64291 Darmstadt, Germany}

\author{K.~Blaum}
\affiliation{Max-Planck-Institut f\"{u}r Kernphysik, Saupfercheckweg 1, 69117 Heidelberg, Germany}

\author{Y.~H.~Zhang}
\affiliation{Institute of Modern Physics, Chinese Academy of Sciences, Lanzhou 730000, People's Republic of China}

\author{Y.~J.~Yuan}
\affiliation{Institute of Modern Physics, Chinese Academy of Sciences, Lanzhou 730000, People's Republic of China}

\author{X.~W.~Ma}
\affiliation{Institute of Modern Physics, Chinese Academy of Sciences, Lanzhou 730000, People's Republic of China}

\author{X.~H.~Zhou}
\email{zxh@impcas.ac.cn}
\affiliation{Institute of Modern Physics, Chinese Academy of Sciences, Lanzhou 730000, People's Republic of China}

\author{H.~S.~Xu}
\affiliation{Institute of Modern Physics, Chinese Academy of Sciences, Lanzhou 730000, People's Republic of China}

\date{\today}

\begin{abstract}
Lifetime measurements of $\beta$-decaying highly charged ions have been performed in the storage ring CSRe by applying the isochronous Schottky mass spectrometry.
The fully ionized $^{49}$Cr and $^{53}$Fe ions were produced in projectile fragmentation of $^{58}$Ni primary beam and were stored in the CSRe tuned into the isochronous ion-optical mode.
The new resonant Schottky detector was applied to monitor
the intensities of stored uncooled $^{49}$Cr$^{24+}$ and $^{53}$Fe$^{26+}$ ions.
The extracted half-lives $T_{1/2}(^{49}{\rm Cr}^{24+})=44.0(27)$~min and $T_{1/2}(^{53}{\rm Fe}^{26+})=8.47(19)$~min
are in excellent agreement with the literature half-life values corrected for the disabled electron capture branchings.
This is an important proof-of-principle step towards realizing the simultaneous mass and lifetime measurements on exotic nuclei
at the future storage ring facilities.
\end{abstract}


\maketitle

\section{Introduction}
The half-life is a fundamental property of an atomic nucleus.
The knowledge of nuclear half-lives is important for understanding nuclear structure and in nuclear astrophysics~\cite{Litvinov11}.
Even though the half-life of a neutral atom is almost independent of external physical and chemical conditions~\cite{Emery72},
the modification of the number of orbital electrons in the atom has a great effect on its half-life~\cite{Litvinov07,Winckler09}.
Furthermore, since atoms are highly ionized in hot stellar environments,
the $\beta$ decay rates of highly charged ions (HCI) are worth being investigated in order to accurately determine the time scales and pathways of stellar nucleosynthesis processes~\cite{Atanasov15}.

To experimentally address half-lives of HCIs, one needs to be able to produce exotic nuclides in the required high atomic charge states and to preserve the ions in these states for sufficiently long time.
The development of heavy-ion cooler storage rings coupled to in-flight fragmentation facilities offers such a possibility.
Highly-charged radioactive ions are produced at relativistic energies in a production target, which simultaneously acts as an electron stripping target.
Owing to the ultra-high vacuum conditions of a storage ring, the produced ions of interest can be stored in the defined atomic state.
The storage time is defined by the radioactive decay on the one side and by the unavoidable beam losses in the ring on the other side.
Apart from first half-life measurements of HCIs in Electron-Beam Ion Traps \cite{Leach17},
extensive investigations of decays of HCIs were done only in the heavy-ion storage ring ESR at GSI in Darmstadt~\cite{Litvinov11,Bosch13}.
There, a non-destructive monitoring of the number of stored ions in the ring is done by employing the technique of time-resolved Schottky Mass Spectrometry (SMS) \cite{Litvinov04}.

Numerous highlight results were obtained and reported over the 25 years of running the ESR such as
the investigations of bound-state beta decay~\cite{Jung92,Bosch95,Ohtsubo05} or ``modulated'' orbital electron capture decays~\cite{Litvinov08, Kienle13}.
All these experiments were performed with electron cooled ion beams.
Besides well-known advantages of electron cooling there are several disadvantages if half-life measurements are considered.
Firstly, the cooling takes time, ranging from a few seconds to a few minutes, depending on the momentum distribution of particles and its offset to the velocity of electrons in the cooler.
To some extend, this can be improved by implementing the stochastic pre-cooling of the particles before applying the electron cooling \cite{Geissel04,Geissel07}.
Secondly, the recombination of stored ions with cooler electrons is the main atomic beam loss mechanism in the storage ring \cite{Litvinov11}.
Last but not least, the electron cooling damps the amplitudes of betatron oscillations, which prevents two different, electron-cooled ions with very close mass-over-charge ratios to pass each other in the ring.
This effect can cause systematic errors in the determination of the number of stored particles \cite{Litvinov05}.

Isochronous mass spectrometry (IMS) was developed at the ESR to address masses of nuclei with half-lives shorter than the cooling time~\cite{Franzke08, Hausmann00, Geissel06}.
No electron cooling is applied in the IMS thus avoiding the disadvantages listed above.
However, only time-of-flight (TOF) detectors \cite{Trotschr92, Bo10}, based on the recording of secondary electrons from thin foils penetrated by the stored ions, could be used in the rings.
Such detectors destroy the stored beam within a few hundred revolutions, corresponding to a storage time of about a millisecond.
Still, the half-lives of short-lived isomeric states could be addressed with IMS with TOF detectors \cite{Sun10a,Sun10b}.
Owing to the development of high-sensitivity non-destructive Schottky detectors \cite{Nolden11, Zang11, Sanjari13},
which are capable to detect single stored heavy highly-charged ions within a few ten milliseconds,
it became attractive to develop an isochronous Schottky mass spectrometry with time-resolved Schottky detection technique for half-life measurements of HCIs.
Furthermore, this is a basis for half-life measurements of exotic nuclides foreseen at the future storage ring projects at FAIR in Darmstadt and HIAF in Huizhou \cite{Walker13, Zhang16}.
We note, that within the later projects further developments of Schottky detectors to include position-sensitivity has been proposed~\cite{Sanjari15,Chen16}.

Schottky mass spectrometry in the isochronously tuned storage ring has been shown to work in the ESR~\cite{Hausmann00, Sun11}.
In this work we present the first application of IMS+SMS technique to half-life measurements of highly-charged radionuclides.

\section{Experiment}
The operation of the Cooler-Storage Ring at the Heavy Ion Research Facility in Lanzhou (HIRFL-CSR)
enables a new opportunity for $\beta$ decay lifetime measurements of HCIs.
The HIRFL-CSR is quite similar to the high-energy part of the GSI facility,
consisting of the heavy-ion synchrotron SIS,
the in-flight fragment separator FRS,
and the experimental storage ring ESR.
At HIRFL-CSR, the main storage ring (CSRm), used as a heavy-ion synchrotron, is connected to an experimental storage ring (CSRe) via a projectile fragment separator (RIBLL2)~\cite{Xia02}.

With these facilities, HCIs like bare, hydrogen-like (H-like) and helium-like (He-like)
ions can be produced by fragmenting primary beams extracted from CSRm at a sufficiently high energy.
The produced fragments are separated by the RIBLL2 and transferred and injected into the CSRe for experiments.
The CSRe has already been running for about ten years~\cite{Tu09}.
Many masses of short-lived nuclei have been precisely measured by applying the IMS technique at ESR and CSRe~\cite{Knobel16, Geissel06, Stadlmann04, Hausmann00, Tu11,Tu11-1,Xu16}.

The experiment reported here was performed in the context of isochronous mass measurement on $^{52}$Co at CSRe,
where RIBLL2 and CSRe were set to a fixed magnetic rigidity of $B \rho = 5.8574$ Tm~\cite{Xu16}.
The details of the standard IMS measurements at CSRe can be found in Refs.~\cite{Tu11, Tu11-1, Xu16}.
The $^{58}$Ni$^{19+}$ beam accumulated and accelerated in CSRm was fast extracted
and focused upon a 15 mm-thick beryllium production target placed at the entrance of RIBLL2.
The beam intensity was about 10$^8$ particles per spill.
According to the CHARGE calculations~\cite{Scheidenberger98}, at this kinetic energy more than 99.9\% of all fragments emerged the target as fully-stripped atoms.
These HCIs were separated by RIBLL2, which was operated as a pure magnetic rigidity analyzer, and then injected into the CSRe.
The fragments produced in the projectile fragmentation reaction show a quite broad velocity distribution.
The nuclides with different mass-to-charge ratios ($m/q$), e.g. nuclides of interest and calibration nuclides having different velocities, can simultaneously be stored in the CSRe.
This allows for an access to the properties of a series of nuclides simultaneously.
Compared to the SMS with electron cooling the mass resolving power is reduced from about 10$^6$ to 10$^5$~\cite{Litvinov10}.
An energy of 430.8 MeV/u for the primary $^{58}$Ni beam
was chosen to optimize the transmission and storage of the $A=2Z+1$ and $A=2Z+2$ nuclides,
where $A$ and $Z$ are the atomic mass and the proton numbers, respectively.

%===============  fig. 1  ========================================
\begin{figure}[h!]
\begin{center}
\includegraphics[width=8cm]{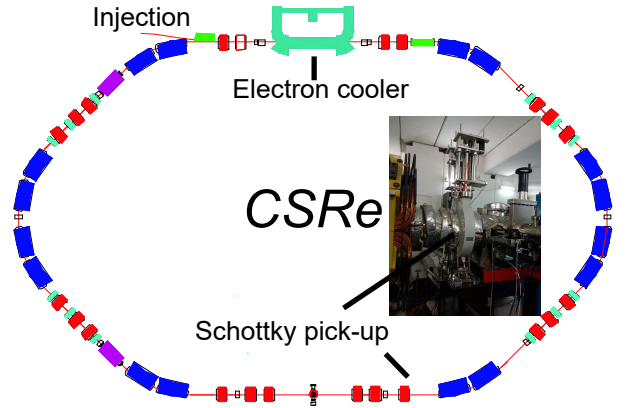}
\caption{The heavy ion storage ring CSRe. The Schottky resonator (see photo in the insert) is installed in a straight section of CSRe. }
\label{fig1}
\end{center}
\end{figure}
%=================================================================
A new, highly-sensitive Schottky resonator was manufactured by GSI \cite{Nolden11}.
It has been installed at one of the straight section of the CSRe~\cite{Wu13}, as indicated in figure~\ref{fig1}.
The principle of the resonator is similar to the one of an antenna.
The stored ions were circulating in the CSRe with revolution frequencies of about 1.6 MHz and at each revolution they excited the resonator.
The signal from the Schottky resonator was analyzed with a commercial real-time spectrum analyzer Tektronix RSA5100A.
Typically the frequencies around the 150$^{\rm th}$ harmonic of the revolution frequency were analyzed.
The IQ (in-phase and quadrature phase) data in the time domain were acquired.
The sampling frequency was set to 3.125~MS/s for IQ.
The RSA5100A was triggered with a period of 5 s by a logical signal. 
Each acquired data file contains in total 409600 samples, corresponding to about 131 milliseconds of recording time.
In order to obtain Schottky frequency spectra for the $A=2Z+1$ and $A=2Z+2$ nuclides simultaneously, a span of 2.5~MHz centered at 243.85~MHz was chosen in the data acquisition.
Duration of each measurement between two adjacent injections into the CSRe was about 30 min.

\section{Data analysis}
\begin{figure}[h!]
\begin{center}
\includegraphics[width=8cm]{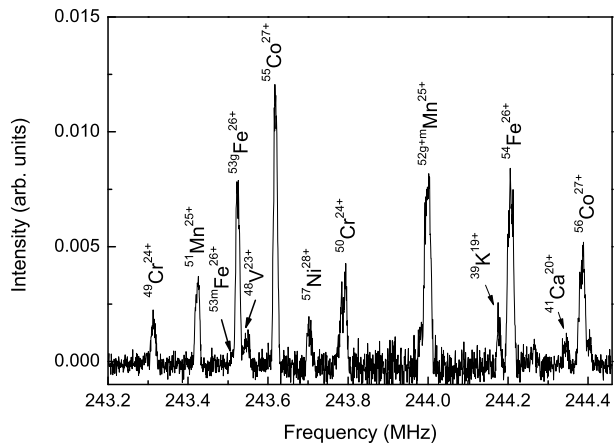}
\caption{Typical Schottky frequency spectrum. The peaks are labelled with the corresponding isotope identification. }
\label{fig2}
\end{center}
\end{figure}

Schottky frequency spectra were obtained by Fast Fourier Transform (FFT) of the IQ data.
4096 IQ data were used to produce one frequency frame.
In order to improve the signal-to-noise characteristics, 600 subsequent frequency frames obtained from 6 subsequently acquired data files were averaged to produce a final frequency spectrum.
One part of the final frequency spectrum is shown in figure~\ref{fig2}.
The resonance response of the Schottky detector was obtained from data where no ions were stored and was subtracted from the spectrum.

%===============  fig. 2  ========================================
%=================================================================

In the isochronous ion-optical mode of the CSRe, the energy of the stored ions of interest is chosen such that $\gamma \approx \gamma_t = 1.4$,
where $\gamma $ is the relativistic Lorentz factor and $\gamma_t$ denotes the transition point of the storage ring \cite{Franzke08}.
The velocity spreads of ions, which are due to the nuclear reaction process, are in first order compensated by the lengths of orbits in the CSRe and can be neglected.
Hence, the revolution frequencies ($f$) reflect directly the $m/q$ ratios of the stored ions, see Ref.~\cite{Franzke08} for more details:
%===============  eq. 1  ========================================
\begin{equation}
\frac{\Delta f}{f}=-\frac{1}{\gamma_t^2} \frac{\Delta
(m/q)}{(m/q)}\quad.
\label{eq1}
\end{equation}
%=================================================================

Particle identification of the frequency peaks in the Schottky frequency spectrum can be done by comparing it with a reference spectrum~\cite{Tu11-1}.
We note, that without cooling, the mean velocities for different ion species are different.
This leads to the situation, see figure~\ref{fig2}, that two different series of nuclides from two different harmonics (154$^{th}$ for $A=2Z+1$, 156$^{th}$ for $A=2Z+2$) are present in the same observation frequency window.

The basic property of Schottky signals is that the area under a frequency peak is proportional to the number of stored ions~\cite{Litvinov11},
%===============  eq. 2  ========================================
\begin{equation}
S\propto2N[\tilde{k}(f)Qef]^2
\quad,
\label{eq2}
\end{equation}
%=================================================================
where $S$ is the area, $N$ and $Q$ are the number and charge state of stored ions, respectively.
The function $\tilde{k}(f)$ represents the sensitivity of the Schottky resonator at different frequencies, $f$.
In our analysis, this factor $\tilde{k}(f)$ is treated as almost constant for a narrow frequency span.
Once $\beta$ decay happens, the $m/q$ ratio changes and the area under the frequency peak will decrease,
which is the cornerstone of the lifetime measurements~\cite{Jung92, Winckler09, Irnich95, Litvinov03}.

The RIBLL2 was operated purely as a magnetic rigidity analyzer.
The advantage is that the nuclides of interest with $A=2Z+1$ and the calibration nuclides with $A=2Z+2$ could be stored simultaneously,
as it is the case for in figure~\ref{fig2}.
The range of mass-to-charge ratios is about 2\%.

\begin{figure}[h!]
\begin{center}
\includegraphics[width=8cm]{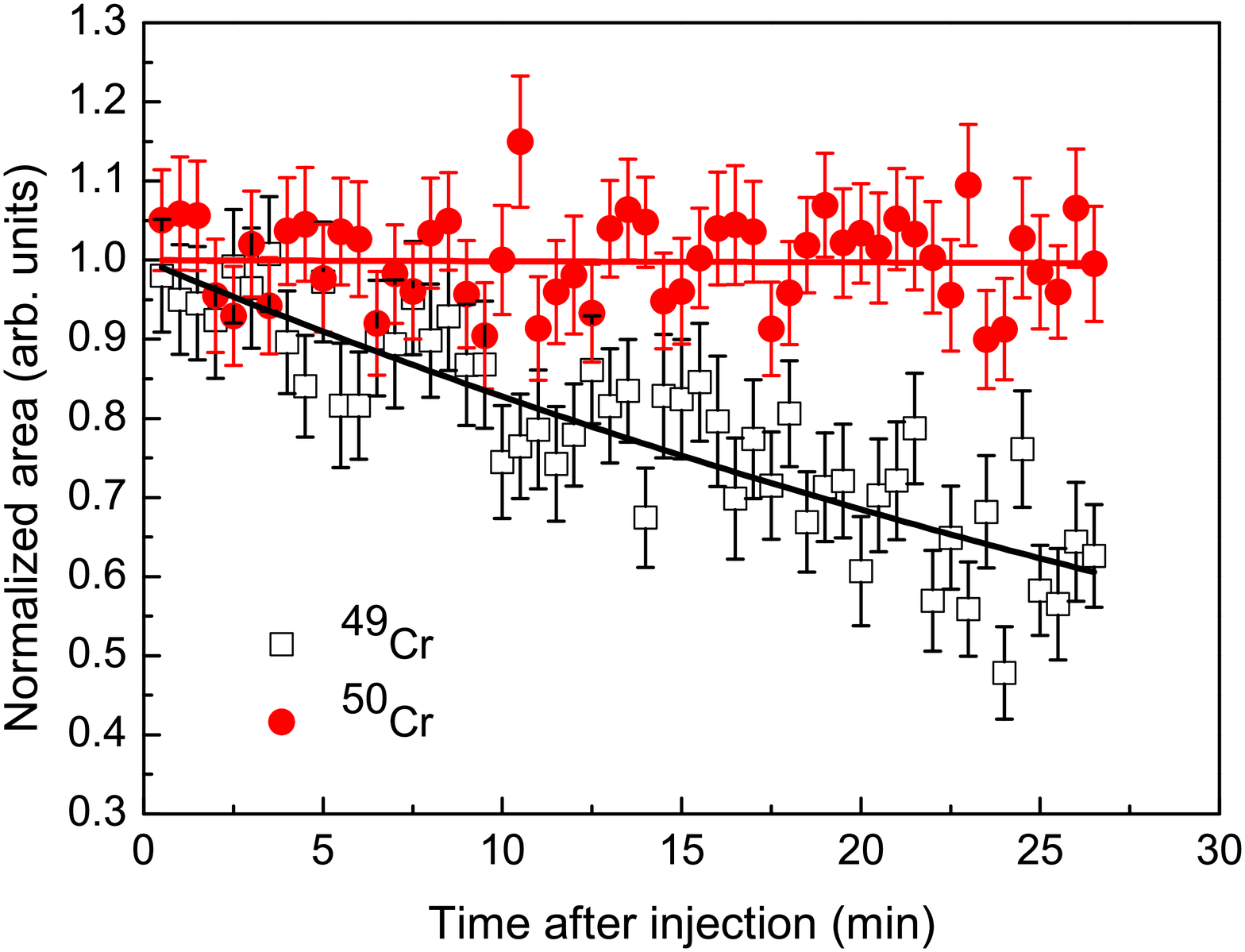}
\includegraphics[width=8cm]{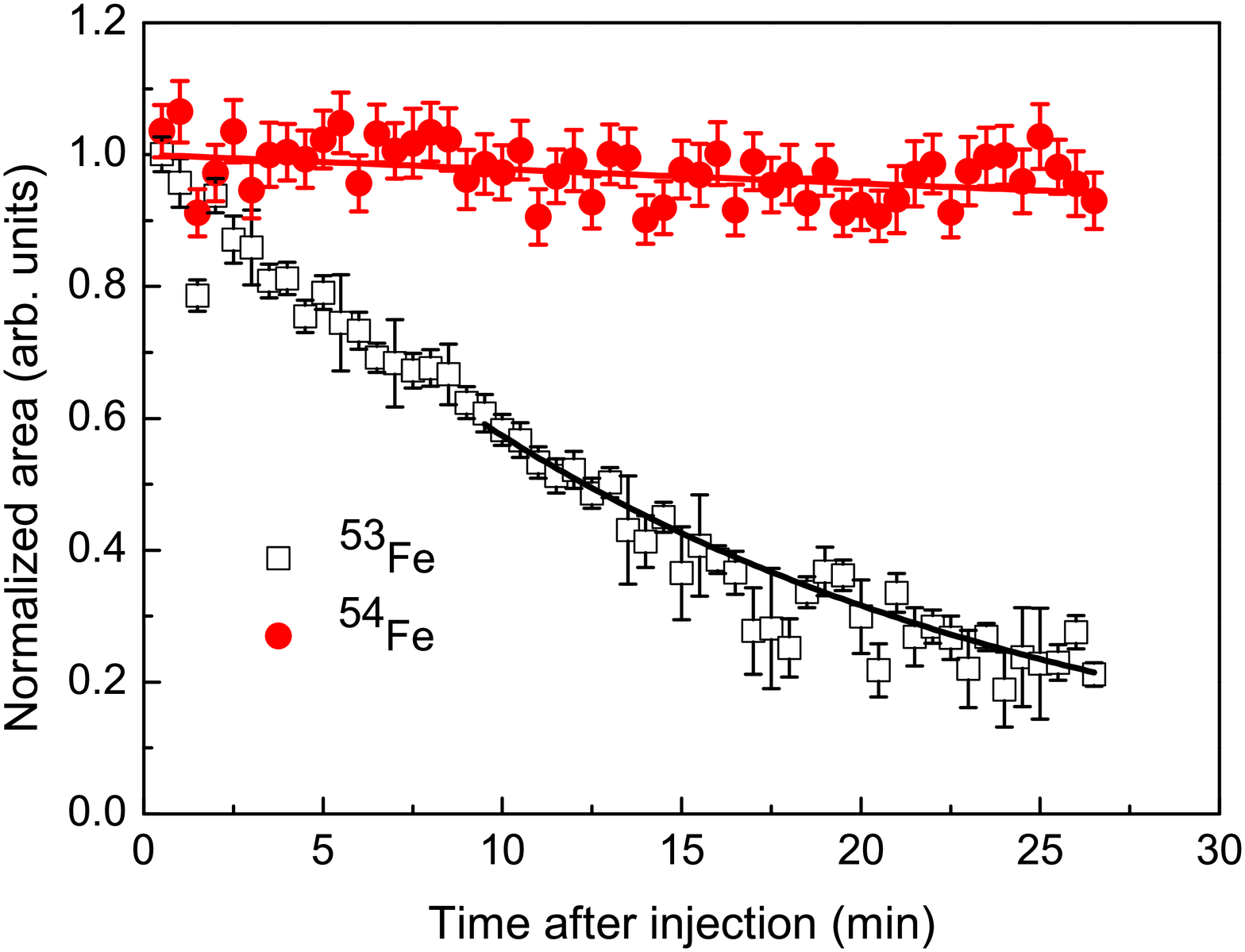}
\caption{Peak areas normalized with the initial area ($S_0$) as a function of the time elapsed since the injection of $^{49}$Cr-$^{50}$Cr and $^{53}$Fe-$^{54}$Fe ions into the CSRe.
The error bars represent statistical uncertainties of the peak areas extracted from the corresponding Gaussian fits.
The solid lines represent fits with single exponential function. The data for $^{53}$Fe is fitted only starting from time $T > 9$ min, when the isomers have mostly decayed. For details see text.}
\label{fig3}
\end{center}
\end{figure}

Taking into account that averaging of 600 frequency frames was done for each frequency spectrum,
the statistical uncertainty of the amplitude at a given frequency point is about 4\%~\cite{Winckler09, Schlitt97}.
The areas under the frequency peaks were extracted by means of Gaussian fitting.
The normalized peak areas of  $^{49}$Cr-$^{50}$Cr and $^{53}$Fe-$^{54}$Fe as a function of the time elapsed since injection are plotted in figure ~\ref{fig3}.

%===============  fig. 3  ========================================
%=================================================================

The $\beta$ decay constant ($\lambda_\beta$) was extracted by fitting the ratios ($R$=$S^{aim}$/$S^{cali.}$) of the peak areas with a single exponential function:

%============================  eq. 3 ================================

\begin{eqnarray}
R&=& \frac{S_{0}^{ aim}exp(-\lambda_tT)}{S_{0}^{cali.}exp(-\lambda_sT)}\nonumber\\
   &=& \frac{S_{0}^{ aim}exp[-(\lambda_s+\lambda_\beta)T]}{S_{0}^{cali.}exp(-\lambda_sT)}\quad,
\label{eq3}
\end{eqnarray}
%=================================================================
where $S^{aim}$ and $S^{cali.}$ are the peak areas of aimed  nuclide and calibration nuclide, respectively.
 $T$ is the time elapsed since injection.
The total decay constant, $\lambda_t$, is the sum of the beam loss constant, $\lambda_s$, and $\beta$ decay constant, $\lambda_\beta$.
The bare $^{50}$Cr and $^{54}$Fe ions are stable, which means that their $\beta$ decay constants  are equal to zero.
The  beam loss constants depend typically only on $Z$ and follow a quadratic form for different nuclei~\cite{Irnich95}.
Please note, that the main loss mechanism in Ref.~\cite{Irnich95} was the recombination in the electron cooler, which is disabled in our case.
The average storage constants obtained for $^{50}$Cr and $^{54}$Fe are 7.2(19)~h and 9.0(26)~h, respectively.
We note, that the mean velocity difference of about 1\% for the nuclides
of interest and the calibration nuclides results in the correspondingly different beam loss constants.
However, taking into account much larger statistical uncertainties of our measured values, such a small difference can be neglected.
The obtained $\beta$-decay half-lives are listed in table~\ref{tab1}.
%========================tab.1=========================================
\begin{table}[h!]
\caption{Measured half-lives of bare $^{49}$Cr and $^{53}$Fe nuclides ($T_{1/2}^{CSRe}={\rm ln}(2)/\lambda_\beta/\gamma$).
Listed are the half-lives of neutral atoms  ($T_{1/2}$), theoretical half-lives of bare ions ($T_{1/2}^{cal}$), and the experimental half-lives of bare ions  ($T_{1/2}^{ESR}$) from previous ESR measurements~\cite{Irnich95}.}
 \begin{center}
\begin{tabular}{lllll}
 \hline\hline
Nucl.           &~~~~$T_{1/2}$/min &~~~~$T_{1/2}^{cal}$/min &~~~~$T_{1/2}^{ESR}$/min&~~~~$T_{1/2}^{CSRe}$/min\\
                    &~~~~Neutral             &~~~~Bare                           &~~~~Bare                             &~~~~Bare                               \\
\hline
 $^{49}$Cr &~~~~42.3(1)              &~~~~45.6(1)                      &~~~~                                     &~~~~44.0(27)                          \\
 $^{53}$Fe &~~~~8.51(2)              &~~~~8.77(2)                      &~~~~8.5(3)                           &~~~~8.47(19)                          \\

\hline\hline
\end{tabular}
\label{tab1}
\end{center}
\end{table}
%=================================================================

The $\lambda_\beta$ of $^{49}$Cr determined in this work is about 0.0116(7) min$^{-1}$ in the laboratory frame.
Taking into account the Lorentz factor $\gamma=1.362$, deduced from the magnetic rigidity of CSRe, $T_{1/2}(^{49}{\rm Cr}^{24+})=44.0(27)$~min in the rest frame.
The half-life for neutral $^{49}$Cr is 42.3(1) min and the $\beta^+$ branching ratio $I_\beta=92.8\%$~\cite{Burrows08}.
Thus, the estimated half-life for bare $^{49}$Cr is 45.6(1) min, assuming that the rest 7.2\% entirely proceed via orbital electron capture decay.

The peak of $^{53}$Fe in the ground state contains feeding from the isomer due to isomeric transition.
The half-life for the bare $^{53}$Fe isomer is 2.48(5) min~\cite{Irnich95}.
Therefore, we fit the data for $^{53}$Fe only starting from time $T > 9$ min, when about 85$\%$ of the isomers have already decayed.
The $\lambda_\beta$ of $^{53}$Fe determined in this work is 0.0601(14) min$^{-1}$ in the laboratory frame.
The Lorentz factor $\gamma$ is 1.363 leading to $T_{1/2}(^{53}{\rm Fe}^{26+})=8.47(19)$~min in the rest frame.
The half-life for neutral $^{53}$Fe is 8.51(2) min and the $\beta^+$ branching ratio $I_\beta=97.04\%$~\cite{Huo99}.
Thus, the half-life of 8.77(2) min is expected for bare $^{53}$Fe.
The experimental half-life for bare $^{53}$Fe measured at GSI is 8.5(3) min~\cite{Irnich95}.

The differences of $\beta$ decay Fermi functions between neutral and bare ions are not considered in the estimations of half-lives~\cite{Irnich95}.
The half-lives obtained in this work are in excellent agreement with both theoretical and available experimental values.
The latter indicates that the IMS+SMS technique can be used to study $\beta$ decay lifetimes of HCIs.

\section{Summary and Outlook}
Beta decay half-lives of highly charged ions were measured in the CSRe storage ring tuned into the isochronous ion-optical mode by applying the non-destructive time-resolved Schottky mass spectrometry technique.
The half-lives of  bare $^{49}$Cr and  $^{53}$Fe nuclei were determined.
The results are in excellent agreement with the literature values corrected for the disabled decay channels involving electrons.
The results demonstrate the capability of such isochronous Schottky mass spectrometry to lifetime studies of HCIs,
which is an important proof-of-principle step towards realizing the simultaneous mass and lifetime measurements on exotic nuclei
at the future storage ring facilities FAIR in Germany and HIAF in China.

Concerning the near-term future, half-life measurements at CSRe will be continued
with the investigation of the orbital electron capture decay of bare, H-like, and He-like  $^{111}$Sn ions~\cite{Litvinov09}.
The half-life for neutral $^{111}$Sn is 35.3(6) min and the $\beta$ branching ratio $I_\beta=30.2\%$~\cite{Blachot09}.
It is expected that, due to the conservation of the total angular momentum of the nucleus-lepton system, the hyperfine ground state of H-like $^{111}$Sn ions can not decay via electron capture.
The highly charged $^{111}$Sn ions have already been produced and stored at the CSRe.
Compared to the results for $^{49}$Cr and $^{53}$Fe from this work, the charge state of $^{111}$Sn is much higher which will cause a better signal-to-noise characteristics of the Schottky signal.
The stable primary beam $^{112}$Sn in relevant atomic charge states will be used to calibrate the beam loss constants.

The authors thank the IMP accelerator team for the excellent support.
This work is supported in part by the CAS Pioneer Hundred Talents Program,
by the CAS Open Research Project of large research infrastructures,
by the NSFC (Grants No. U1232208, U1432125, 11605248, 11605249, 11605252, 11605267 and 11775273),
by the Max-Plank-Society,
by the National Key Program for S\&T Research and Development (Grant 2016YFA0400504),
by the Major State Basic Research Development Program of China (Contract No.2013CB834401),
by the Key Research Program of Frontier Sciences, CAS (Grant No. QYZDJ-SSW-S),
by the Helmholtz-CAS Joint Research Group HCJRG-108,
by the External Cooperation Program of the CAS Grant No. GJHZ1305, and
by the ExtreMe Matter Institute EMMI at the GSI Helmholtzzentrum f\"{u}r Schwerionenforschung, Darmstadt, Germany.

\end{document}